\documentclass[11pt]{article}

\usepackage[left=2cm, right=2.5cm, top=2.5cm, bottom=2.5cm]{geometry}
\usepackage[x11names]{xcolor}
\usepackage{amssymb, amsmath, graphicx, pgfplots, tikz,cancel}

\usepackage[colorlinks=true, linkcolor=blue, citecolor=blue, hyperindex=true, linktocpage=true]{hyperref}

\title{\textbf{Lagrangian mechanics on Lie groups: a pedagogical approach}}
\author{Andrew Lucas \\ 
\textsl{Department of Physics, Stanford University} \\ \texttt{ajlucas@stanford.edu}}
\date{\today}

\begin{document}
\maketitle 

\begin{abstract}
We describe a new method to formulate classical Lagrangian mechanics on a finite-dimensional Lie group.  This new approach is much more pedagogical than many previous treatments of the subject, and it directly introduces students to generator matrices and their usefulness in many manipulations.  The example of rigid body rotation, i.e. motion on the Lie group $\mathrm{SO}(3)$, is used as an example, and it is shown how to derive Euler's equations directly from the principle of least action.  The techniques covered in this paper generalize to other Lie groups in a straightforward manner, which is discussed.    We briefly discuss the Hamiltonian formulation of the problem as well.   
\end{abstract}

\tableofcontents

\thispagestyle{empty}

\pagebreak

\setcounter{page}{1}

\section{Introduction}
In this paper, we derive the equations of motion for a classical system whose configuration space is a Lie group, using the Lagrangian and Hamiltonian formalisms.  These are perhaps the most interesting examples of classical dynamics on non-trivial manifolds.   Lie groups are manifolds which, despite behavior which is often topologically nontrivial at a global level, have quite simple properties locally, and can be completely described by a simple set of commutation relations.   They appear repeatedly in diverse branches of physics: for example, the Lie group $\mathrm{U}(1)\times\mathrm{SU}(2)\times\mathrm{SU}(3)$ is the gauge group of the Standard Model \cite{peskin}.  In classical mechanics, the special orthogonal groups $\mathrm{SO}(n)$ relate to $n$-dimensional rigid body rotation \cite{arnold}.

While the equations of motion for systems such as rigid rotators on Lie groups are derived in many mechanics books as \cite{arnold, landau, goldstein} , the derivations avoid the formalism of the principle of least action entirely.   This misses a wonderful opportunity to introduce students to Lie groups in a context which is familiar:  one can certainly pick up rigid bodies and throw them to observe dynamics.  Furthermore, the derivations in traditional textbooks usually do not emphasize that concepts such as angular velocities and momenta take on much more subtle meanings in higher dimensions; thinking of them as vectors on the same footing as translational velocities and positions provides students with bad habits and intuition for much of physics, which increasingly takes place on manifolds beyond $\mathbb{R}^3$.   For example, in  relativity, angular momentum must be described as a tensor \cite{goldstein, landau2}.

The treatment of the present paper provides matrix methods for using Lagrangian and Hamiltonian formalisms on Lie groups, which are nice examples of manifolds where a set of coordinates is not inherently obvious.   This not only provides pleasing theoretical closure to the theory, but it also emphasizes the physical nature of each of the quantities involved in the calculations.  It also provides what is most likely the simplest example of a system where the ability to re-choose local coordinates at each point in time is essential.  Despite these things, the set-up of the problem and the general procedure is similar to simpler problems in Lagrangian mechanics that are thoroughly studied.

We first give a brief introduction to why rigid body rotation is a problem of mechanics on a Lie group.  Then, the paper explores the Lagrangian formulation of mechanics on a Lie group, followed by a brief foray into the Hamiltonian formulation.  Throughout the paper, the specific example of the group $\mathrm{SO}(3)$ is used, as this is the most relevant historically.   Appendices are provided briefly discussing manifolds in mechanics, Lagrangian mechanics on a manifold, and Lie groups, and are necessary background for understanding this work.

The presentation is intended to provide a sketch of how the material could be presented in a graduate classical mechanics course, with most derivations of interest provided directly and in a unified framework.   The reader not familiar with manifolds or Lie groups, including the adjoint representation and generator matrices, is encouraged to first read the appendices, or consult a book on the subject: e.g., the discussion in \cite{peskin}.

\section{Rigid Body Rotation}
Here we present a proof of what is frequently called Chasles' Theorem, which states that the configuration space of a $n$-dimensional rigid body is $\mathbb{R}^n\times \mathrm{SO}(n)$: physically, this means translation and rotation, respectively.   There exist many proofs of this theorem, particularly for $n=3$, although most rely on geometric considerations \cite{arnold}.  Following the theme of this paper, here we use a very direct, computational argument.   While the paper is more general than simply a discussion of rigid body rotation, as both the most practical and famous example of motion on a Lie group, it is appropriate to begin the discussion here.

In general, the displacement of a solid is described by $u(x,t)$, with $u\in \mathbb{R}^n$, $x\in \mathbb{R}^n$ being positions and $t\in \mathbb{R}$ being time.  However, for a rigid solid, the constraints on $u(x,t)$ are strict: all lengths in the solid must be preserved with time, meaning that for any $t_1,t_2 \in \mathbb{R}$ and $x_1,x_2\in \mathbb{R}^n$, \begin{equation}
| u(x_1,t_1)-u(x_2,t_1)| = |u(x_1,t_2)-u(x_2,t_2)|.
\end{equation}
One way to keep this constraint is to simply translate the solid: i.e., $u(x,t_2) = u(x,t_1) + a$.   So  pick some point in the solid (call it $x=0$) to ``follow" in time, and measure all displacements relative to that point: i.e., $u(0,t) \equiv 0$.   Furthermore, because as of now we are free to label the points in the solid, let us define the points so that $u(x,0)\equiv x$, and thus $u$ characterizes displacements from the initial position.

Then we are left with the constraint that \begin{equation}
u^{\mathrm{T}}(x,t)u(x,t) = x^{\mathrm{T}} x.\end{equation}
We can write \begin{equation}
u(x,t) = A(x,t)x
\end{equation}where $A\in \mathbb{R}^{n\times n}$ is a matrix.   This implies that \begin{equation}
x^{\mathrm{T}} (A^{\mathrm{T}}(x,t)A(x,t) - 1)x = 0 \label{chasleseq}.
\end{equation}$A(x,t)$ must be $x$-independent:   if $A$ had an $x$-dependence, then we could take two $x$-derivatives of (\ref{chasleseq}), and the left term in parentheses would be $x$-dependent while the right term would be constant, thus violating the equality that the two must equal.   Note that because $A$ is the same for all points, there are only a finite number of degrees of freedom.

We define $\mathrm{O}(n)$, the orthogonal group, as \begin{equation}
\mathrm{O}(n) = \lbrace B\in\mathbb{R}^{n\times n} \; | \; B^{\mathrm{T}} B = 1\rbrace.
\end{equation} Thinking if $\mathrm{O}(n)$ as a manifold, we see $\mathrm{O}(n)$ has two disconnected parts: one with determinant 1 and one with determinant $-1$.  Because the dynamics of $A$ should be continuous, and we start, by construction, with $A(x,0)=1$, we know that $A\in \mathrm{SO}(n)$, the determinant 1 subgroup of $\mathrm{O}(n)$: \begin{equation}
\mathrm{SO}(n) = \lbrace B\in\mathbb{R}^{n\times n} \; | \; B^{\mathrm{T}} B = 1,\; \det B = 1\rbrace.
\end{equation}   Because $\mathrm{SO}(n)$ is also a manifold, and also has the mathematical structure of a group, because it is closed under matrix multiplication, it is a Lie group.

This concludes the proof of Chasles' theorem:  the configuration space consists of both the translational part and the subsequent rotation part, and is thus $\mathbb{R}^n \times \mathrm{SO}(n)$.

For the remainder of the paper, we ignore the translation on $\mathbb{R}^n$, as this is a frequently-solved problem.   We focus on the dynamics on the Lie group, which is the topic of the present paper.

\section{Lagrangian Formalism}
We now describe how to derive equations of motion on a generic Lie group directly from the principle of least action.  While certainly there are many different possible equations of motion on the same Lie group, the purpose of this discussion however is to introduce a new computational methodology that may be useful for many of them.   We do this by presenting a derivation of the equations of motion of a free rigid rotator on $\mathrm{SO}(n)$ which naturally generalizes: we formally comment on why the generalization makes sense at the end.   

The general procedure is the standard one:  constructing a Lagrangian and then finding the equations of motion (Lagrange's equations); the only difference is that the procedure is kept coordinate-independent until the very end.
\subsection{Constructing a Lagrangian}
In this section we formulate the Lagrangian of a system whose configuration space is $\mathrm{SO}(n)$.   For simplicity, we assume that the system is not in an external potential, as this adds an unnecessary complication.  

Suppose we have a rigid body in $n$ dimensions, rotating about the origin, with mass density $\rho(x)$.   If there are no external forces, the Lagrangian for the continuum motion of the body is given by the kinetic energy, \begin{equation}
L = \int \frac{1}{2}\rho(x) \dot{u}^{\mathrm{T}}(x,t) \dot{u}(x,t) \mathrm{d}^n x.
\end{equation} We have used here the notation from the previous section.   Letting $A \in \mathrm{SO}(n)$ be a matrix describing the orientation of the body, we can rearrange terms and, because \begin{equation}
\dot{u}(x,t) = \dot{A} x,
\end{equation} conclude that \begin{equation}
L = \frac{1}{2} \mathrm{tr}[K\dot{A}^{\mathrm{T}} \dot{A}]
\end{equation}where we have defined the matrix \begin{equation}
K \equiv \int \rho(x) xx^{\mathrm{T}} \mathrm{d}^nx .
\end{equation}
As of now, the Lagrangian is not expressed in terms of a coordinate chart, but this actually is helpful.  For example, the Lagrangian certainly must have a global $\mathrm{SO}(n)$ symmetry in the sense that if we simply rotated our frame of reference, we should not change the dynamics.   That is manifest in the expression above, because a transformation $A\rightarrow BA$ for $B$ a time independent $\mathrm{SO}(n)$ matrix certainly leaves $L$ invariant.   Furthermore, if we keep the Lagrangian in coordinate independent terms until we actually solve Lagrange's equations, we can pick local coordinates which both make the calculation more simple, elegant and easy to manipulate, and serve to emphasize the physics. 
\subsection{Lagrange's Equations}
Now, we turn to the methods which can be used to solve Lagrange's equations.  We repeatedly comment upon the connections with the mathematics to physical intuition.

Firstly, we note that we expect that because of the global $\mathrm{SO}(n)$ symmetry, the equations of motion should only depend on the velocities in whatever generalized coordinates we choose.   Secondly, we note that locally, we can write any matrix in $ \mathrm{SO}(n)$ as $\exp[\alpha^a T^a] A_0$ for small $\alpha^a$.   So, if we want the equations of motion at time $t=t_0$, we can pick coordinates so that $A_0 = A(t_0)$.   This means that $\alpha^a(t_0)=0$.   $\alpha^a$ will serve as the set of generalized coordinates for the motion at $t_0$.   One of the first major points of this discussion is therefore that we will exploit our freedom to re-define the $\alpha^a$ at each point in time.   Note that we have made no comment as to what the generators $T^a$ are, or what $A_0$ is: we'll come back to this point shortly.   Physically, think of this as the observer rotating his coordinate axes slightly at each time step, so that his coordinate axes are oriented in the same way relative to the body at each time.

Given the methodology, let's express $L$ in terms of these generalized coordinates.   To do so, we use the fact that \begin{align}
\dot{A}(t_0) &= \left.\frac{\mathrm{d}}{\mathrm{d}t}\left(\mathrm{e}^{\alpha^a(t) T^a} A_0\right)\right|_{t=t_0} = \left.\left( \dot{\alpha}^aT^a A_0+ \mathrm{O}(\alpha)\right)\right|_{t=t_0} = \dot{\alpha}^a T^a A_0  \label{timederivative}
\end{align}where we have used the fact that since the $\alpha^a$ vanish at $t=t_0$, only the first term contributes.   (In general, we would have higher order corrections involving multiple generators.)   This implies that, taking the transpose of the above equation, using the antisymmetry of the generators and the cyclic property of the trace: \begin{equation}
L = - \frac{1}{2}\mathrm{tr}[A_0K{A_0}^{\mathrm{T}} T^b T^c] \dot{\alpha}^b \dot{\alpha}^c. 
\end{equation}Let's define the matrix \begin{equation}
M^{ab} = -\frac{1}{2}\mathrm{tr}[A_0 K{A_0}^{\mathrm{T}} \lbrace T^a,T^b\rbrace]
\end{equation} where $\lbrace T^a,T^b\rbrace = T^a T^b + T^b T^a$ is the anticommutator.   This matrix will show up frequently in the equations below.   We can then write \begin{equation}
L = \frac{1}{2}M^{bc} \dot{\alpha}^b \dot{\alpha}^c. \label{genleg}
\end{equation}Evidently $M^{ab}$ is like a mass/moment of inertia tensor.

Obtaining Lagrange's equations must be done with care, but the procedure emphasizes many mathematical aspects which are irrelevant to the discussion on a trivial manifold where one coordinate chart covers the entire manifold.   We begin by computing $\partial L/\partial \alpha^a$.   This derivative is in general non-zero, because the matrix $M^{ab}$ subtly carries an $\alpha^a$ depedence in it through the $A_0$.   Using the fact that (by the reasoning of  (\ref{timederivative})) \begin{equation}
\frac{\partial A_0}{\partial \alpha^a} = T^a A_0,
\end{equation}we obtain that using the cyclic property of the trace, \begin{equation}
\frac{\partial L}{\partial \alpha^a} = \frac{1}{4}\mathrm{tr}(A_0 K {A_0}^{\mathrm{T}} [T^a, \lbrace T^b,T^c\rbrace]) \dot{\alpha}^b\dot{\alpha}^c.
\end{equation}Since \begin{align}
[T^a, T^bT^c] &= T^b[T^a, T^c] + [T^a,T^b]T^c = f^{acd}T^b T^d + f^{abd}T^d T^c,
\end{align}we can simplify: \begin{align}
\frac{\partial L}{\partial \alpha^a} &= \frac{1}{2}\mathrm{tr}(A_0 K{A_0}^{\mathrm{T}} \lbrace T^b, T^d\rbrace) f^{acd} \dot{\alpha}^b \dot{\alpha}^c = f^{adc} M^{bd}\dot{\alpha}^b \dot{\alpha}^c \label{coordder}
\end{align}  

Now, we compute the canonical momenta: \begin{equation}
p^a \equiv \frac{\partial L}{\partial \dot{\alpha}^a} = M^{ab}\dot{\alpha}^b.
\end{equation}The time derivative here is deceptively simple: \begin{equation}
\frac{\mathrm{d}p^a}{\mathrm{d}t} = M^{ab}\ddot{\alpha}^b.   \label{lgtd}
\end{equation}The reason is subtle but extremely important.    Because the momentum $p^a$ is only well-defined at a precise point on the manifold, and the coordinate chart is specifically changed at each point on the manifold, we are forced to use the coordinate chart at one point.  At this point, $M^{ab}$ is a constant, dependent only on the position on $\mathrm{SO}(n)$.  

Putting together (\ref{lgtd}) and (\ref{corder}), we find \begin{equation}
M^{ab} \dot{\omega}^b = f^{adc}M^{db} \omega^b\omega^c \label{euler}
\end{equation}where $\omega^a\equiv \dot{\alpha}^a$.  These are the generic equations of motion on a $\mathrm{SO}(n)$; indeed, their form is the same on other Lie groups, if the Lagrangian is quadratic in velocities.

\subsection{Symmetries and Conservation Laws}
Before connecting the discussion to rigid body rotation on $\mathrm{SO}(3)$, let's investigate some consequences of Noether's theorem.  Certainly the energy is conserved since $\partial L/\partial t = 0$, but here we are interested in consequences of group symmetries.   Finding conserved quantities is always a good thing, because they allow us to reduce the number of differential equations that must be solved.

As per usual, if there is any $\alpha^a$ for which (for some choice of generators) $\partial L/\partial \alpha^a = 0$, then $p^a$ is a conserved quantity.  By looking at  (\ref{coordder}) this will happen for generic velocities when \begin{equation}
0= f^{adc} M^{db} + f^{adb} M^{dc}.   \label{consmomentum}
\end{equation} for fixed $a$, and all $b$, $c$.   We discuss the physical meaning of this in the next subsection.

There is a more interesting class of symmetries, which come only from the group structure.  If we define $p= p^a T^a$, $\mathrm{tr}(p^k)$ is an independent constant of motion for even $k$ which are less than $n$, for $\mathrm{SO}(n)$ dynamics  \cite{dickey}.  Let us prove this for $k=2$.   For any (semisimple) Lie group we can choose generators so that $\mathrm{tr}(T^a T^b) \sim \delta^{ab}$, $\mathrm{tr}(p^2) = \mathrm{tr}(T^a T^b) p^a p^b \sim p^a p^a$.   From the equations of motion, \begin{equation}
\frac{\mathrm{d}(p^ap^a)}{\mathrm{d}t} = 2p^a \dot{p}^a = 2M^{ae}\dot{\alpha}^e f^{adc}M^{bd} \dot{\alpha}^b\dot{\alpha}^c.
\end{equation}
Since $M^{ab} = M^{ba}$ and we can work in any basis, let us choose one in which $M^{ab}$ is diagonal (this can always be done, as one can show with linear algebra).  This implies that \begin{equation}
\frac{\mathrm{d}(p^ap^a)}{\mathrm{d}t} = 2 f^{abc} M^{aa} M^{bb} \dot{\alpha}^a \dot{\alpha}^b \dot{\alpha}^c = 0,
\end{equation}where to obtain the last equality we used the fact that $f^{abc} = -f^{bac}$.   We have used the Einstein summation convention whenever there are two or more indices of the same letter.  In particular, we have shown that on a generic Lie group, the square of the ``angular momentum vector" is a constant of motion.

\subsection{The Case of SO(3)}
Let us now turn to the case of $\mathrm{SO}(3)$.   This corresponds to the motion of a free rigid body (no applied external forces) in $n=3$ dimensions, and is the classic example of motion on a Lie group.  We now exploit the fairly generic results from before, pointing out the physical meaning of the various quantities.  

First, note that the generators of $\mathrm{SO}(n)$ are antisymmetric matrices in $\mathbb{R}^{3\times 3}$.  A reminder of why is included in the appendix.   This means that the generators may be written in the form \begin{subequations}\begin{align}
T^1 &= \left(\begin{array}{ccc} 0 &\ 0 &\ 0 \\ 0 &\ 0 &\ 1 \\ 0 &\ -1 &\ 0 \end{array}\right),  \\
T^2 &= \left(\begin{array}{ccc} 0 &\ 0 &\ -1 \\ 0 &\ 0 &\ 0 \\ 1 &\ 0 &\ 0 \end{array}\right),  \\
T^3 &= \left(\begin{array}{ccc} 0 &\ 1 &\ 0 \\ -1 &\ 0 &\ 0 \\ 0 &\ 0 &\ 0 \end{array}\right).
\end{align}\end{subequations}
By computing commutators, we determine that \begin{equation}
[T^a,T^b] = \epsilon^{abc}T^c,
\end{equation}so we conclude that the structure constants $f^{abc}$ for this group are no different than the Levi-Civita tensor in 3 dimensions, $\epsilon^{abc}$.

Further manipulations with these generator matrices lead to \begin{subequations} \begin{align}
\lbrace T^1, T^1\rbrace &= \left(\begin{array}{ccc} 0 &\ 0 &\ 0 \\ 0 &\ -2 &\ 0 \\ 0 &\ 0 &\ -2 \end{array}\right), \\
\lbrace T^1, T^2\rbrace &= \left(\begin{array}{ccc} 0 &\ 1 &\ 0 \\ 1 &\ 0 &\ 0 \\ 0 &\ 0 &\ 0 \end{array}\right).
\end{align}\end{subequations} and various cyclic permutations of the indices and respective rows/columns.  This implies that, e.g., \begin{subequations}\begin{align}
M^{11} &= \int \rho(x)\left((x^2)^2 + (x^3)^2\right) \mathrm{d}^3x , \\
M^{12} &= -\int \rho(x) x^1 x^2 \mathrm{d}^3x,
\end{align}\end{subequations}and cyclic permutations of indices, in agreement with well-known results for the $\mathrm{SO}(3)$ rigid rotator.   Indeed, $M^{ab}$ is the moment of inertia tensor.   This derivation leads to important insights: quantities such as angular velocity and momentum, and tensors such as the moment of inertia, have indices in the adjoint representation.   While for $\mathrm{SO}(3)$, the fundamental and adjoint representations have the same dimension (and are basically the same), this is not true in general.

From here, we finally use the freedom to choose $A_0$.  If we choose $A_0$ to make $M$ diagonal with entries $M^1$, $M^2$ and $M^3$,  (\ref{euler}) becomes\begin{equation}
M^1 \dot{\omega}^1 = \omega^2 \omega^3 (M^2-M^3)
\end{equation} and cyclic permutations of indices, in agreement with the historical results.    

We also comment on the case when $\partial L/\partial \alpha^1=0$, on a chart where $M^{ab}$ is diagonal.    (\ref{consmomentum}) says that \begin{equation}
f^{abc} M^{bb} + f^{acb}M^{cc} = \epsilon^{abc}(M^{bb} - M^{cc})  = M^2 - M^3 =0,
\end{equation}or that $M^2=M^3$.   This is also an important historical result, although it is usually presented in the other way (e.g., in \cite{goldstein}): if two of the eigenvalues of $M^{ab}$ are equal, there is a conserved component of the angular momentum, which is in this case $p^1$.

It is worthwhile to explain to students that the cross product of two vectors $v$ and $w$ in $\mathbb{R}^3$ can be thought of as follows:   denote $v =v^a e^a$, $w=w^ae^a$ where $e^a$ are the standard basis vectors for $\mathbb{R}^3$.   If we instead think of $V = v^a T^a$ and $W = w^a T^a$, where now these vectors are antisymmetric matrices, then we can think of $\mathbf{v}\times\mathbf{w}$ as equivalent to $[V,W]$.   This follows directly from the commutation relations of the generator matrices.    We see that the cross product is ultimately quite special and unique to 3 dimensional spaces, in allowing us to map matrix commutation to a type of vector multiplication.

Before moving on, it may be worthwhile to inform students that there is another Lie group with the same algebra as $\mathrm{SO}(3)$: $\mathrm{SU}(2)$:\begin{equation}
\mathrm{SU}(2) \equiv \lbrace B\in\mathbb{C}^{2\times 2}\; |\; B^\dagger B = 1, \; \det B = 1\rbrace.
\end{equation}These groups, it turns out, are not isomorphic, meaning there is not a unique correspondence between an entry of $\mathrm{SO}(3)$ and an entry of $\mathrm{SU}(2)$:   in $\mathrm{SU}(2)$, rotations by $2\pi$ degrees take one to $-1$, whereas in $\mathrm{SO}(3)$ these rotations take one back to the identity.   This distinction is meaningful, as in quantum mechanics spinors (of spin $\frac{1}{2}$) transform under $\mathrm{SU}(2)$ rotations, leading to the nontrivial property that linear combinations of spinors are not always left invariant under rotations by $2\pi$ degrees. \cite{sakurai}

\subsection{Generalization to Other Lie Groups}
Let us briefly discuss why the form of the equations of motion found in this section are the equation of motion for ``free particle" motion on any Lie group.     The lowest order Lagrangian which is nontrivial will be quadratic in the velocities, as in  (\ref{genleg}).   The coordinate dependence of $L$ is embedded in $M^{bc}$.  Now, the canonical momenta and their derivatives are the same as before because $L$ is the same.  But we also expect $\partial L/\partial \alpha^a$ to be identical.   This is because $M^{bc}$ is a tensor with indices in the adjoint representation, and we expect that each of them will transform in this representation under an adjoint representation ``rotation": \begin{align}
\delta M^{bc} = \delta \alpha^a \left[ f^{abd} M^{dc} + f^{acd} M^{bd} \right] + \mathrm{O}(\delta \alpha^2).
\end{align} This equation implies that we should expect for $\partial M^{bc}/\partial \alpha^a$, and thus $\partial L/\partial \alpha^a$, to behave exactly as it did earlier.   This argument provides a mathematical justification for the equations of motion, whereas the previous one provided a more physical interpreration.

It is also worth mentioning that if $M^{ab} = M \delta^{ab}$: i.e., the particle is a ``point mass" of mass $M$ which lives on a Lie group, then the equation of motion can also be thought of as the geodesic equation \cite{dowker}.   While a modification of this argument likely would also lead to a derivation of the equation of motion in the case of $M^{ab}$ not proportional to the identity, such an approach is far less intuitive to students who are just learning about Lie groups, as it requires associating the Christoffel coefficients, familiar from general relativity, with the structure constants.
\section{Hamiltonian Formalism}
We briefly comment on this formalism in the context of Hamiltonian mechanics.  The Hamiltonian formalism of mechanics on manifolds is the study of symplectic geometry.  Because there are many sources on this subject such as \cite{arnold, ana}, and the level of mathematics is significantly higher, we will mostly neglect a discussion of the Hamiltonian formalism.   We will briefly describe it in the context of a Legendre transform from Lagrangian formalism.   One benefit of the Hamiltonian formalism is that it is never unclear when to take a derivative of a quantity such as $M^{ab}$, simply because the equations of motion are first order.

Let $(M^{-1})^{ab}$ be the matrix such that $(M^{-1})^{ab}M^{bc} = \delta^{ac}$.   Using the canonical momenta $p^a$ found earlier, a simple Legendre transform of \begin{equation}
H = p^a \dot{\alpha}^a - L
\end{equation}leads to \begin{equation}
H = \frac{1}{2}(M^{-1})^{ab}p^a p^b.
\end{equation}  As in the Lagrangian case, the $M^{-1}$ hides the dependence on the local coordinate chart $\alpha^a$.   Nonetheless, using the identity $(M^{-1})^{ab} M^{bc} = \delta^{ac}$ and the effective calculation of $\partial M^{ab}/\partial \alpha^c$ as done in  (\ref{coordder}), we obtain Hamilton's equations to be \begin{subequations}\begin{align}
\frac{\partial H}{\partial p^a} &= (M^{-1})^{ab}p^b = \dot{\alpha}^a, \\
-\frac{\partial H}{\partial \alpha^a} &= f^{abc}(M^{-1})^{cd} p^b p^d  = \dot{p}^a.   \label{momeq}
\end{align}\end{subequations} The latter equation is readily seen to be identical to the Euler equations found previously, as expected.

\section{Conclusion}
In summary, this paper presented a derivation of generalized Euler's equations for a Lie group.  While this has been accomplished by authors in the past, the present work provides needed insight by presenting a  computational trick, which is both efficient and pedagogical, to determine the equations of motion directly from the principle of least action.  The chosen Lagrangian can thus be justified on either physical or mathematical grounds.   The presentation is also much more appropriate for graduate students in physicis without graduate-level mathematics backgrounds in geometry; much of the treatment of mechanics on a Lie group is in the mathematical literature and is not presented in a form easily accessible by students of classical mechanics.

The formalism provided in this paper is a bit more advanced than the usual discussions on rigid body motion: however, it also provides students with an introduction to Lie theory in the more familiar setting of finite-dimensional classical mechanics.   A good understanding of Lie groups and their representations is crucial to understanding modern particle physics, and many physicists, such as the author,  never see Lie theory until a study of Yang-Mills gauge theories in quantum field theory.  Introducing the mathematics sooner would help to familiarize students more with the techniques in a field of physics where they are more comfortable.

Of course, we also believe that this paper provides satisfactory methods for tackling mechanics on Lie groups, manifolds whose coordinate charts are challenging to obtain directly.  In particular, the derivation of Euler's equations from first principles is far more satisfactory than typical derivations, and provides deep physical insight about how the nontrivial nature of the manifold leads to the nonlinear terms in the equations.  

\addcontentsline{toc}{section}{Acknowledgements}
\section*{Acknowledgements}
The author would like to thank D. Anninos, K. Luna and S. Raghu for helpful comments, suggestions and discussions.
\appendix
\section{Manifolds and Mechanics}
The configuration space of a system is the set of all possible states or orientations of our physical system, at any given instant in time, neglecting the rate of change of the system.   We call that set $Q$; in classical mechanics, we always take $Q$ to be a manifold; see \cite{arnold} for details on the precise definition of a manifold.  The exact details are unimportant: think of them as spaces on which one can do calculus (there is a notion of smoothness) and which locally look like Euclidean spaces $\mathbb{R}^k$ for some integer $k$, but which globally may look completely different.

Suppose that we have a free point particle of mass $m$ moving along an infinite  line in 1 dimension.  In this case, $Q=\mathbb{R}$, the set of real numbers.   For our purposes, the manifold is trivial, because we only need to define one coordinate, $x$, to describe every point on $Q$.   We say that we have defined a coordinate chart, which provides an invertible map from (at least some subset of) $Q$ to (a subset of) $\mathbb{R}^k$ (where $k$ is the dimension of the manifold); in this case, $k=1$.

Let's consider another manifold:  $Q=\mathrm{S}^1$, or the circle (or 1-dimensional sphere, which we usually think of as embedded in a 2-dimensional space).   If we just look at the circle in some tiny region, it still looks like a line; however, if we look at it ``globally" it is distinct from the line:  it closes back on itself, and is topologically distinct from $\mathbb{R}$.  The circle can be parameterized by an angular coordinate $\theta$, as is typically done in physics.  However, you still only need to define $\theta$ once to describe every point on the manifold uniquely.   A common map sends $\mathrm{S}^1$ to the subset $[0,2\pi)$ of $\mathbb{R}$.

Now take $Q=\mathrm{S}^2$: the 2-dimensional sphere (usually thought of as embedded in a 3-dimensional space).    Once again, this manifold locally appears flat:  think of standing on the Earth and looking at its surface:   it is hard to tell that it is curved.   A typical coordinate chart here consists of angular coordinates $(\theta,\phi)$ (take $\phi$ as the azimuthal coordinate).   This time, however, we run into problems.    If $\theta = 0$ or $\pi$, then the chart to $\mathbb{R}^2$ is no longer invertible:  any choice of $\phi$ works just as well.    This is not always some abstract problem that is not important in practice.   If one writes down the Lagrangian for a particle moving along the sphere, the equations of motion will fail at $\theta=0$.    In fact, it can be proven that there is no choice of coordinates that avoids this problem: at least one point will always be excluded in any chart from $\mathrm{S}^2$ to $\mathbb{R}^2$.   The obvious intuition is, therefore, to simply change the coordinates and re-orient them so that the particle avoids $\theta=0$.   In fact, this is a useful trick in Lagrangian and Hamiltonian mechanics:  one can easily change the coordinate chart if the equations of motion will become problematic, and it is one we will exploit for good effect in deriving equations on a Lie group.

It is sometimes the case in physics that $Q$ is a manifold so abstract it is hard to visualize directly.   We prove in this paper that $Q=\mathrm{SO}(n)$ for a $n$-dimensional rigid body with one point fixed.   $\mathrm{SO}(n)$ is a subset of the set of all $n\times n$ real-valued matrices, $\mathbb{R}^{n\times n}$ which is both a group and a manifold.   Even for $n=3$, the case of interest, $\mathrm{SO}(3)$ is a 3-dimensional space that has many non-trivial properties as a manifold, the most important of which, for our sake, is that it takes multiple coordinate charts to cover $\mathrm{SO}(3)$.   

While this paper is not intended to be a primer in topology, and very little topology is in fact needed to understand the results of this paper, it is important for students to understand how nontrivial manifolds arise as configuration spaces for systems that appear frequently in problems and in real life.  The rigid body provides an excellent example of a practical problem whose configuration space is too abstract to even picture geometrically.

\section{Lagrangian Mechanics on a Manifold}
For convenience here, as in the body of the paper, we assume that the Lagrangian is time-indepdendent.   See  \cite{arnold} for more detail on many of the points below.

Let $Q$ be a smooth manifold, and denote by $\mathcal{C}^\infty(Q)$ the set of smooth functions $f:\mathbb{R}\rightarrow Q$.   We define the action $S$, a functional on $Q$, as a map $S:\mathcal{C}^\infty (Q) \rightarrow \mathbb{R}$ with the property that trajectories which make $S$ an extremum are the physical trajectories realized by the system.  This principle of least action is useful because we also assume we can write the action as an integral in time over the Lagrangian $L:\mathrm{T}Q\rightarrow \mathbb{R}$: \begin{equation}
S[q(t)] = \int\limits_{t_1}^{t_2} L(q(t), \dot{q}(t)) \mathrm{d}t.
\end{equation}Here $t_1$ and $t_2$ are the starting and ending times, respectively, and $\mathrm{T}Q$ is the tangent bundle of the manifold $Q$, or the set of both points on $Q$ and tangent vectors (velocities) at those points.

The standard Lagrange's equations are recovered, and can be derived in the usual way, by choosing a coordinate chart locally on $Q$.   If this local coordinate chart is given by $\lbrace q^i\rbrace$ (where the index $i$ takes on a different value for each degree of freedom), we can express Lagrange's equations as \begin{equation}
\frac{\mathrm{d}}{\mathrm{d}t} \frac{\partial L}{\partial \dot{q}^i} = \frac{\partial L}{\partial q_i}.
\end{equation}This is a famous result, but we take special care to analyze it here.  In particular, we remember that $\dot{q}\in \mathrm{T}_qQ$, the tangent fiber to the manifold at point $q$.  This is a fancy word for the subset/submanifold of points in $\mathrm{T}Q$ that correspond to position $q$, and arbitrary velocity $\dot{q}$.   When we do variational calculus in the action to derive Lagrange's equation, we are actually evaluating $\partial L/\partial \dot{q}_i$ at the point $q+\delta q$, not $q$.   Indeed, for a generic manifold, it makes no sense to talk about a tangent vector at point $q_1$ as tangent at point $q_2$, even if those points are very close to each other.  On a Lie group this seemingly pedantic point has very real consequences.

Having said all of this about manifolds and Lagrangian mechanics on them, it is worth emphasizing to students how little actually changes.  The procedure of Lagrangian mechanics is identical; the only thing one should be careful of is that choosing a coordinate chart can be tricky, and that exploiting the freedom to re-choose coordinate charts (at each time) can greatly simplify the solution to the problem.

\section{Lie Groups}
A Lie group $G$ is a manifold endowed with a group multiplication operation which is smooth  \cite{ana}.   We will represent the elements of $G$ in this paper by square real or complex matrices, so that the group operation is simply matrix multiplication.   For a similarly brief presentation intended for physicists, see \cite{peskin}.

Because $G$ is a manifold and a group, it contains the identity matrix as well as elements arbitrary close to the identity.   We can write these elements $g\in G$ as \begin{equation}
g = 1 + \epsilon^a T^a + \mathrm{O}(\epsilon^2)
\end{equation} for infinitesimal $\epsilon^a$; in this equation, 1 is the identity matrix, $\lbrace T^a\rbrace$ is a set of matrices called generators for the group, and we employ the Einstein summation convention.   Note that the choice of each $T^a$ is not unique; changing $T^a$ will just require changing $\epsilon^a$ to leave $\epsilon^a T^a$ invariant.   If $\dim G = n$, the index $a$ will have $n$ possible values.   

As an example, the Lie group $\mathrm{SO}(n)$ can be defined as \begin{equation}
\mathrm{SO}(n) = \lbrace g \in \mathbb{R}^{n\times n} \; | \; \det g = 1, \; g^{\mathrm{T}} g = 1\rbrace.
\end{equation} We can find the generators of $\mathrm{SO}(n)$ as follows:  since we must maintain $g^{\mathrm{T}} g = 1$, we know that \begin{align}
(1+\epsilon^a T^a)^{\mathrm{T}} (1+\epsilon^a T^a) &= 1 + \epsilon^a (T^a + (T^a)^{\mathrm{T}}) + \mathrm{O}(\epsilon^2) \notag \\
&= 1 
\end{align}
which implies that \begin{equation}
(T^a)^{\mathrm{T}} = -T^a,
\end{equation}i.e. the generators of $\mathrm{SO}(n)$ are antisymmetric $n\times n$ matrices.   Since there are $\frac{1}{2}n(n-1)$ linearly independent antisymmetric $n\times n$ real-valued matrices, we know that $\dim (\mathrm{SO}(n)) = \frac{1}{2}n(n-1)$.

For the purposes of this paper, it will be convenient to look at the Lie groups as slightly more abstract objects than as just matrix groups.  In general, we can think of a Lie algebra for the generators, given by the commutators of the generators.  If we define structure constants $f^{abc}$ so that \begin{equation}
[T^a, T^b] = f^{abc}T^c
\end{equation} then we can define the Lie group by the $f^{abc}$ instead.   The commutation relations above are also called the Lie algebra of the given Lie group.  It can be shown that we can always choose $f^{abc}$ to be totally antisymmetric in its indices.  A representation of the the Lie group refers to a set of matrices which can be written as $\exp[\alpha^a T^a]$ for some $\alpha^a$, where $\lbrace T^a\rbrace$ are a set of generators which obey the Lie algebra.   In this paper, we will often find that the physically intuitive coordinates for describing (a subcomponent of) the system will be in the fundamental representation of the Lie group.  This means that, in the case of $\mathrm{SO}(n)$, these objects are vectors in $\mathbb{R}^n$; e.g., these are the coordinates of a point in a $n$-dimensional rigid body.   We will primarily be focused on the adjoint representation.  If $d$ is the dimension of the group, the generators are defined as the $d\times d$ matrices with entries given by \begin{equation}
(T^a_{\mathrm{Adj}})^{bc} = f^{abc}
\end{equation}in the adjoint representation.   We will always use a set of coordinates $\alpha^a$ in the adjoint representation to describe dynamics on a Lie group.

\addcontentsline{toc}{section}{References}
\bibliographystyle{plain}
\bibliography{mymechbib}

\end{document}